\documentclass{PoSmod}
\DeclareSymbolFont{letters}{OML}{txmi}{m}{it}

\usepackage{url}
\usepackage{mciteplus}
\usepackage{amsbsy}   
\usepackage{amsfonts}
\usepackage{amsmath}
\usepackage{amssymb}

\newcommand{\BR}[2]{\mathrm{BR}\mathinner{(#1\rightarrow #2)}}

\newcommand{\FH}{{\tt FeynHiggs}}
\newcommand{\THDMC}{{\tt 2HDMC}}
\newcommand{\SuperIso}{{\tt SuperIso}}

\newcommand{\HB}{{\tt HiggsBounds}}

\title{Tools for charged Higgs bosons}

\ShortTitle{Tools for charged Higgs bosons}

\author{\speaker{Oscar St{\aa}l}\\
        Deutsches Elektronen-Synchrotron DESY\\
        Notkestra{\ss}e 85\\
        D-22607 Hamburg, Germany\\
        E-mail: \email{oscar.stal@desy.de}}

\ReportNo{DESY 10-236 }

\abstract{We review the status of publicly available software tools applicable to charged Higgs physics. A selection of codes are highlighted in more detail, focusing on new developments that have taken place since the previous charged Higgs workshop in 2008. We conclude that phenomenologists now have the tools ready to face the LHC data. A new webpage collecting charged Higgs resources is presented.}

\FullConference{Third International Workshop on Prospects for Charged Higgs Discovery at Colliders - CHARGED2010,\\
		September 27-30, 2010\\
		Uppsala Sweden}

\begin{document}

\section{Introduction}
The ability to make and use tools is essential to mankind. It is necessary for our survival and well-being. In fact, it is so important that  anthropologists have claimed it to be a defining characteristic of our species \cite{ToolMaker}. Surely, our high standing technological society would be unthinkable were it not for the plethora of tools available for different purposes. Needless to say, the level of Higgs physics would not be very advanced either. 

The LHC experiments will hopefully become the ultimate tools to study charged Higgs bosons. Until this is reality, we rely on theoretical tools to make predictions. Theorist's tools are usually computer codes which can be applied to calculate some interesting observables from model input. Sometimes such tools are made available to the public, and we should all feel grateous towards those investing their time and careers in this effort. Going from a private code to a public release often means a deviation from the straight path to publication and instant fame. Instead it leads into an endless cycle of bug fixing, improving user interfaces, writing manuals, and the occasional glorious moment of releasing a new version on the web. If the relase is successful, people start using the program, which leads to user feedback and the author can go back to fix the new bugs and restart the cycle. 

\section{Toolbox for charged Higgs physics}
Charged Higgs bosons---which we generically denote by $H^\pm$---appear in any non-trivial extension of the Standard Model (SM) Higgs sector.\footnote{Non-trivial refering to the transformation properties of the new scalar field under $\mathrm{SU}(2)_L$. Also a scalar $\mathrm{SU}(2)$ singlet carrying non-zero hypercharge  leads to a charged Higgs boson, as realized in the Zee model \cite{Zee:1980ai}.} This is interesting, since the presence of a charged scalar is something fundamentally different; there is no SM particle with the same quantum numbers. Doublets have a special position among the possible representations with renormalizable couplings to the SM, since they do not upset the the tree-level relation $\rho=M_W/M_Z\cos\theta_W\simeq 1$. Guided by the principle of parsimony, most studies are performed on models with two Higgs doublets (2HDM). Another strong argument in favor of the the 2HDM is of course that this model is the minimal Higgs sector compatible with supersymmetry (SUSY).

In this note we will only discuss tools which are publicly available. Since we are not aware of any codes dealing with exotica (e.g.~charged SU(2) singlets or Higgs triplet models), the scope will be limited to the (SUSY and non-SUSY) 2HDM. There are many calculations concerning charged Higgs bosons which tie into more general problems, such as computing the SUSY particle spectrum at the electroweak scale from GUT-scale parameter input. A full coverage clearly goes beyond what can be discussed here. Instead we present a fairly complete list of tools for different aspect of charged Higgs physics at the URL 
\begin{itemize}
\centering
\item[] \url{http://www.desy.de/~stal/chtools}
\end{itemize}
We aim to maintain this list and keep it up to date. If you have a code which is related to charged Higgs boson physics, and it is not in this list, we are more than willing to add it. Please contact the author directly.

\section{Focus on a few selected tools}
\subsection{FeynHiggs}
A fundamental task in Higgs physics is to compute the masses and couplings of the Higgs bosons. The leading program for doing these calculations in the MSSM since many years is \FH\ \cite{Heinemeyer:1998yj}. Two other alternatives are {\tt HDecay} \cite{Djouadi:1997yw} and {\tt CPSuperH} \cite{Lee:2003nta}. Among other things, \FH\ gives the most accurate predicition for the charged Higgs mass $m_{H^\pm}$ available (when not used as an input parameter).
At tree-level, $m_{H^\pm}$ is related to the CP-odd Higgs boson mass $m_A$ through
\begin{equation}
m_{H^\pm}^2=m_A^2+m_W^2.
\label{eq:mhpma}
\end{equation}
Unlike the corrections to the lightest CP-even Higgs mass $m_h$---which are often sizable---the mass relation \eqref{eq:mhpma} typically receives only moderate corrections at the one-loop level. Nevertheless, these corrections are important to achieve the precision required to compare with the ultimate sensitivity of the LHC (and later the linear collider) \cite{SvenTalk}.
 In the Feynman-diagrammatic approach, the one-loop corrected $m_{H^\pm}$ is given by the pole of the charged Higgs propagator, obtained by solving the equation
\begin{equation}
q^2-m_{H^\pm}^2+\Sigma_{H^+H^-}^{(1)}(q^2)=0,
\end{equation}
where $\Sigma_{H^+H^-}$ is the charged Higgs self-energy. The calculations in \FH\ allow for both real and complex parameters \cite{Frank:2006yh}. The latter is a prerequisite for treating CP violation in the Higgs sector, something which is forbidden at tree-level in the MSSM, but which can be induced by loop effects. For the neutral Higgs masses and mixing matrices, the full one-loop corrections are included, and also the known two-loop corrections.  At the two-loop level, corrections to Equation~(\ref{eq:mhpma}) proportional to $\mathcal{O}(\alpha_s y_t^2)$ are known in the approximation where the electroweak gauge couplings are set to $g=g'=0$.
 
\FH\ also calculates the charged (and neutral) Higgs decay modes, including leading QCD corrections. Another important class of corrections which are included are the non-holomorphic corrections to the $b$-quark Yukawa coupling (so-called $\Delta_b$-corrections). These affect the $tbH^\pm$ coupling and can lead to substantial suppression (or enhancement) of the branching ratio for $H^\pm\to tb$ and to the production cross section at hadron colliders \cite{Hashemi:2008ma}. \FH\ contains many additional features, such as the calculation of flavor observables, corrections to $m_W$, $g-2$ for the muon, and parametrized LHC cross sections for both neutral and charged Higgs production.

\subsection{2HDMC -- Two-Higgs-Doublet Model Calculator}
The two-Higgs-doublet model calculator (\THDMC) \cite{Eriksson:2009ws} is a fairly new code, on which work was initiated as a direct result of the cHarged 2008 workshop. It can be used to perform calculations in a general (not necessarily supersymmetric) version of the 2HDM. This model is described by the Higgs potential
\begin{equation}
  \begin{aligned}
    V_{\rm{2HDM}} = &\,m_{11}^2\Phi_1^\dagger\Phi_1+m_{22}^2\Phi_2^\dagger\Phi_2
    -\left[m_{12}^2\Phi_1^\dagger\Phi_2+\mathrm{h.c.}\right]
    \\
    &+\frac{1}{2}\lambda_1\left(\Phi_1^\dagger\Phi_1\right)^2
    +\frac{1}{2}\lambda_2\left(\Phi_2^\dagger\Phi_2\right)^2
    +\lambda_3\left(\Phi_1^\dagger\Phi_1\right)\left(\Phi_2^\dagger\Phi_2\right)
    +\lambda_4\left(\Phi_1^\dagger\Phi_2\right)\left(\Phi_2^\dagger\Phi_1\right)
    \\&+\left\{
    \frac{1}{2}\lambda_5\left(\Phi_1^\dagger\Phi_2\right)^2
    +\left[\lambda_6\left(\Phi_1^\dagger\Phi_1\right)
      +\lambda_7\left(\Phi_2^\dagger\Phi_2\right)
      \right]\left(\Phi_1^\dagger\Phi_2\right)
    +\mathrm{h.c.}\right\},
  \end{aligned}
  \label{eq:pot_gen}
\end{equation}
with two identical scalar doublets $\Phi_1$, $\Phi_2$. The parameters $m_{12}^2$ and $\lambda_5$--$\lambda_7$ can be complex, while the remaining parameters are real. Assuming CP conservation, it is possible to find a basis in which \emph{all} parameters are real. This is the case currently treated by \THDMC. Following electroweak symmetry breaking, there are in total eight free parameters (compared to two in the MSSM). These can be specified using different parametrizations, e.g.~the physical Higgs masses. Higgs masses and mixings are computed at tree-level. Note that the ratio of the two vacuum expectation values, $\tan\beta\equiv v_2/v_1$ at this stage is \emph{not} a physical parameter since the potential is invariant under rotations in the Higgs space.

A full phenomenological specification of the 2HDM requires also the Yukawa couplings, which are of the general form
\begin{equation}
\mathcal{L}_{\mathrm{Yuk}}=\overline{Q}_LY^U_i\widetilde{\Phi}_i U_R+\overline{Q}_LY^D_i\Phi_i D_R+\overline{L}_LY^U_i\Phi_i E_R+\mathrm{h.c.},
\end{equation}
where a sum over $i=1,2$ is implied. Only one linear combination of each set of Yukawa matrices $Y^F_i$ (corresponding to the fermion mass matrix $M^F$) can be diagonalized. The orthogonal combination---which governs the coupling of the charged Higgs boson---can only be simultaneously diagonal under the assumption of some symmetry relation among the $Y_i$. Most commonly a $Z_2$ symmetry is used to implement the Glashow-Weinberg criterion \cite{Glashow:1976nt}. The resulting Yukawa sectors are known as 2HDM `types'. Another option is the so-called aligned model \cite{Pich:2009sp}, where a linear relation $Y^F_1=\xi^F Y_2^F$ is imposed. In \THDMC\ the Yukawa sector can be specified using any of these prescriptions---or in a completely free fashion---which offers a great deal of flexibility in which models can be studied.

\begin{figure}
\centering
\includegraphics[width=0.45\columnwidth]{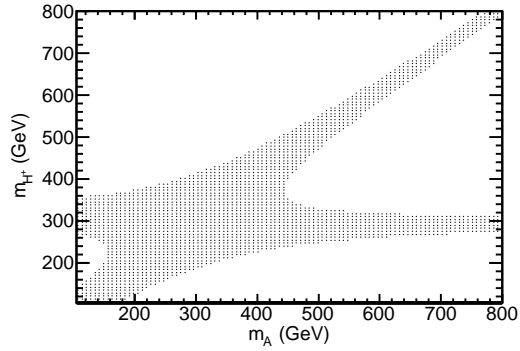}
\caption{Constraints on $m_{H^\pm}$ in the general 2HDM from the oblique $T$ parameter. The shaded region is allowed at $2\,\sigma$. The CP-even Higgs masses are $m_h=117$~GeV, $m_H=300$~GeV, and $\sin^2(\beta-\alpha)=1$.}
\label{fig:T}
\end{figure}
In addition to the Higgs spectrum, \THDMC\ can be applied to calculate theoretical constraints on the 2HDM from positivity and unitarity, it computes the Higgs decay modes (including QCD corrections where applicable and some off-shell effects), and the 2HDM contributions to the oblique EW parameters. An example of how the latter can be used is shown in Figure~\ref{fig:T}, which shows the constraints on the splitting between $m_{H^\pm}$ and the other `heavy' Higgs masses from the $T$ parameter (using the experimental value $T=0.07\pm0.08$ \cite{Amsler:2008zzb}). The two custodial limits $m_{H^\pm}=m_A$ and $m_{H^\pm}=m_H$ are clearly visible. 

\subsection{HiggsBounds}
The program \HB\ \cite{Bechtle:2008jh} answers a frequently asked question in Higgs phenomenology: is this model excluded by present collider limits? Even if the question is simple, to give the right answer is not. Before the arrival of \HB, most theorists therefore either applied the SM limit $m_h\gtrsim 114$~GeV, or performed a one-dimensional analysis of the coupling $g^2_{hZZ}$ which controls $e^+e^-\to Zh$ production at LEP, to judge the validity of their models. Both approaches are often dubious, and with \HB\  available it is no longer `beyond the scope' to check the collider Higgs mass limits. The code has already been linked with a number of other programs -- including \THDMC presented above. For quick testing of only a few models, a web interface is also available.

\HB\ contains a large collection of results from a number of experimental analyses at LEP and the Tevatron. Any model with $n$ neutral and $m$ charged Higgs bosons can be tested, but the user has to supply the (reduced) couplings and Higgs boson widths (branching ratios). To ensure the correct statistical interpretation as exclusion at $95\%$ CL in the presence of many channels, the model prediction is not compared to \emph{all} analyses; only to the one deemed most sensitive judging by the \emph{expected} exclusion $\eta_{\mathrm{exp}}=\frac{\sigma_{\mathrm{model}}}{\sigma^{95}_{\mathrm{exp}}}$. This single channel is then tested for exclusion by evaluating the ratio of the prediction to the \emph{observed} limit, $
\eta_{\mathrm{obs}}=\frac{\sigma_{\mathrm{model}}}{\sigma^{95}_{\mathrm{obs}}}$. Models with $\eta_{obs}>1$ are excluded.

\begin{figure}
\centering
\includegraphics[width=0.5\columnwidth]{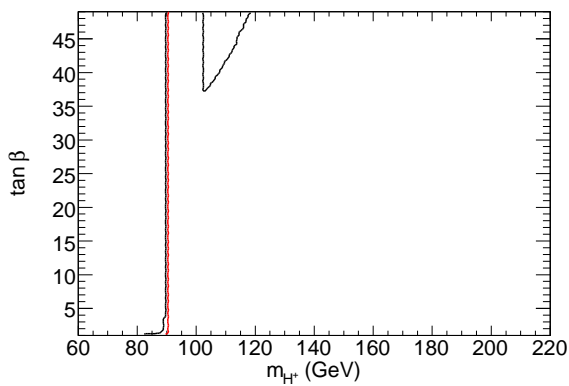}
\caption{Exlusion limits at $95\%$ CL on $m_{H^\pm}$ and $\tan\beta$ from direct $H^\pm$ searches. The results were obtained using \HB\ 2.0.0 linked to \THDMC. Values below $m_{H^\pm}\lesssim 90$ GeV are excluded by LEP, and inside the triangular region by the Tevatron.}
\label{fig:chlim}
\end{figure}

In the latest version (2.0.0), \HB\ includes for the first time limits from direct searches for the charged Higgs boson. Figure \ref{fig:chlim} presents the exclusion limits in a general 2HDM type II, including only the experimental searches for $H^\pm$. As can be seen from the figure, LEP established a firm limit $m_{H^\pm}\gtrsim 90$~GeV, while the Tevatron excludes a fairly small mass range in the high $\tan\beta$ region. With the LHC coming up to steam, these charged Higgs mass limits are expected to improve significantly in the near future. It would be extremely useful to have \HB\ continuously updated with the latest results.

\subsection{SuperIso}
The charged Higgs bosons are not only searched for at high-energy colliders, but they can also play a major role in low-energy processes. In particular for several observables measured in $B$-meson decays, which are sensitive to new charged currents with enhanced couplings to the third generation fermions. In the MSSM with R-parity conservation, $H^\pm$ gives the only new contribution to flavor changing processes at tree-level, e.g. the leptonic decays of pseudoscalar mesons. This can lead to strong and generic constraints on $m_{H^\pm}$ in this wide class of models. To compare a given scenario to the experimental results, the new physics contributions to the releveant observables must be evaluated as a function of the model parameters. This is the purpose of the \SuperIso\ code \cite{Mahmoudi:2008tp} which computes e.g.~$B\to X_s\gamma$, $B_s\to \mu^+\mu^-$, $B_u\to \tau\nu$, and many additional decay modes of $B$, $D$, and $K$ mesons that are of interest.

One feature of \SuperIso\ which makes it easily extendable to new models is that it does not compute the particle spectra internally, but leaves this task to specialized external codes such as {\tt SoftSUSY} \cite{Allanach:2001kg} (for the MSSM), {\tt NMSSMTools} \cite{Ellwanger:2004xm} (NMSSM), and \THDMC\ (general 2HDM). To exchange data between the programs, extensive use is made of the {\tt SLHA} \cite{Skands:2003cj}. In the future, a similar role is expected to be played by its flavor counterpart, the {\tt FLHA} \cite{Mahmoudi:2010iz}.
\begin{figure}
\centering
\includegraphics[width=0.45\columnwidth]{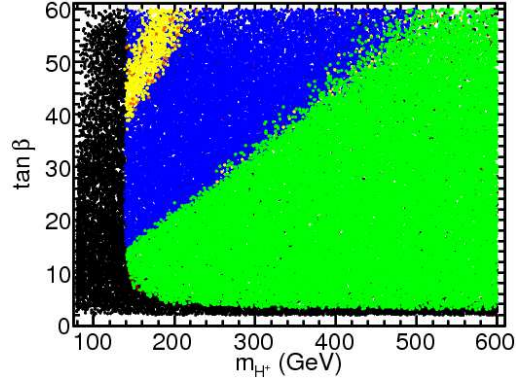}
\caption{Constraints from flavour physics on $(m_{H^\pm},\tan\beta)$ in a scan over SUSY models with non-universal Higgs mass parameters at the GUT scale. The 95\% CL allowed points (green) are plotted on top of exclusion by LEP (black), $B\to\tau\nu$ (blue), $B_s\to \mu^+\mu^-$ (yellow), and $B\to D\tau\nu$ (orange). The figure is taken from \cite{Eriksson:2008cx}.}
\label{fig:flav}
\end{figure}

\SuperIso\ has already been applied to obtain constraints on the properties of charged Higgs bosons in the MSSM \cite{Eriksson:2008cx}, and the 2HDM with general diagonal Yukawa couplings \cite{Mahmoudi:2009zx}. Figure~\ref{fig:flav} comes from the first of these two references. It shows the combined flavor constraints in the $(m_{H^\pm},\tan\beta)$ plane for GUT-based models with non-universal Higgs mass parameters at the unification scale. Note the large exclusion by $B\to \tau\nu$ decays, which are mediated by $H^\pm$ at tree-level.

\subsection{MC@NLO for charged Higgs production}
The description of $H^\pm$ production at hadron colliders is traditionally separated into two different regimes. The \emph{light} charged Higgs ($m_{H^+}<m_t-m_b$), which can be descibed as on-shell $t\bar{t}$ production followed by the decay $t\to bH^+$ ($\bar{t}\to bH^-$). The narrow width approximation is applicable, and the production cross section can be written as the product $\sigma(pp\to t\bar{t})\times\BR{t}{bH^+}$. Several NLO  implementations exist for $\sigma(pp\to t\bar{t})$ \cite{Frixione:2003ei}, and to accurately calculate $\BR{t}{bH^+}$ many tools are available (e.g. \FH\ in the MSSM).\footnote{To obtain a reliable prediction for $\BR{t}{bH^+}$ in SUSY models at high $\tan\beta$, it is essential to include the $\Delta_b$ corrections to the $tbH^+$ coupling.}

When $H^\pm$ is instead heavier than the top quark, there will no longer be an intermediate on-shell top quark. There will instead be associated production of $tH^\pm$, which can be described either in a five-flavor scheme (5FS) as due to $bg\to tH^\pm$, or as $gg\to tbH^+$ (4FS). The two processes require proper matching \cite{Moretti:1999bw}, for which the Monte Carlo (MC) implementation {\tt MATCHIG} \cite{Alwall:2004xw} is available.
The QCD corrections to $tH^\pm$ production are known both in the 5FS \cite{Plehn:2002vy}, and the 4FS \cite{Dittmaier:2009np}. The 5FS calculation was recently implemented \cite{Weydert:2009vr} in the MC@NLO framework \cite{Frixione:2002ik}. 

To have an MC@NLO implementation of this process has several benefits over leading order: the NLO computation offers a reliable normalization of the cross section, a much reduced dependence on the unphysical renormalization and factorization scales, and an `exact' matrix element description of one additional parton. On the other hand, the parton shower approach of the Monte Carlo assures the correct description in the soft and collinear regions of phase space. Unlike a pure fixed order partonic calculation, the implementation in an event generator also has the advantage of producing dressed events (with hadronization, underlying event, etc.) which are fully exclusive and ready for detector simulation. We strongly encourage the experimental collborations to implement the use of MC@NLO for all further charged Higgs analyses.

\section{Summary and conclusions}
Software tools are essential to particle physics. We have introduced the toolbox for charged Higgs physics, and highlighted the physics aspects of some of the tools it contains in more detail. There has been rapid development of tools for charged Higgs physics since the previous workshop in 2008.
To summarize, I would like to emphasize in particular three recent achievements:
\begin{itemize}
\item The MC@NLO code for $tH^\pm$ production, which was actually on the wishlist already from cHarged 2006, has been completed. This is the first MC@NLO implementation of a new physics process that is part of the official release.
\item \HB, which makes the comparison of model predictions to vast amounts of experimental data on Higgs exclusion fast and simple. It makes it into this list especially since it now includes limits from charged Higgs searches.
\item \THDMC, which covers most phenomenological aspects of the general (CP-conserving) 2HDM. We hope that the existence of this code can lead to increased activity and further the collaboration between theory and experiment on the exploration of these models. 
\end{itemize}
Naturally, there has also been continued development and improvements on most other tools during the last years. We think Higgs phenomenology in general is well-equipped to meet the LHC data. Of course, a few interesting areas of possible development were identified and discussed in the course of cHarged 2010. This ensures that not only LHC discoveries---but also some interesting new tools---will be reported on in two years from now.

\section*{Acknowledgments}
I would like to thank the organizers of cHarged 2010 for the invitation to present this talk, and for hosting such a nice and stimulating workshop. 

\bibliographystyle{JHEP}
\bibliography{os_charged10}

\end{document}